# Continuous nucleation switching driven by spin-orbit torques

C. H. Wan[1], M. E. Stebliy[2], X. Wang[1], G. Q. Yu[1,5], X. F. Han[1,3,5*], A. G. Kolesnikov[2], M. A. Bazrov[2], M. E. Letushev[2], A. V. Ognev[2], A. S. Samardak[2,4**]

1. Beijing National Laboratory for Condensed Matter Physics, Institute of Physics, University of Chinese Academy of Sciences, Chinese Academy of Sciences, Beijing 100190, China
2. School of Natural Sciences, Far Eastern Federal University, 690950, Vladivostok, Russia
3. Center of Materials Science and Optoelectronics Engineering, University of Chinese Academy of Sciences, Beijing 100049, China
4. National Research South Ural State University, Chelyabinsk 454080, Russia
5. Songshan Lake Materials Laboratory, Dongguan, Guangdong 523808, China

**Abstract**: Continuous switching driven by spin-orbit torque (SOT) is preferred to realize neuromorphic computing in a spintronic manner. Here we have applied focused ion beam (FIB) to selectively illuminate patterned regions in a Pt/Co/MgO strip with perpendicular magnetic anisotropy (PMA), soften the illuminated areas and realize the continuous switching by a SOT-driven nucleation process. It is found that a large in-plane field is a benefit to reduce the nucleation barrier, increase the number of nucleated domains and intermediate states during the switching progress, and finally flatten the switching curve. We proposed a phenomenological model for descripting the current dependence of magnetization and the dependence of the number of nucleation domains on the applied current and magnetic field. This study can thus promote the birth of SOT devices, which are promising in neuromorphic computing architectures.

Magnetization switching driven by spin-orbit torques (SOT) can be potentially applied in magnetic random-access memory (MRAM) [1-10], spin logics [11-17] and spintronic neuromorphic computing [18-20]. For the MRAM and spin logic applications, sharp magnetization switching between two binary states (for example, spin-up and spin-down states) is preferred, which favors a fast and high on/off ratio for the digital devices. Contrariwise, continuous switching with many accessible intermediate states is more suitable for neuromorphic computing architectures because a neuron network needs its memory components capable of storing continuously variable connection weights for higher adaptability [21-22].

In principle, magnetization switching can be accomplished in three modes, (i) coherent switching and incoherent switching dominated by (ii) domain wall motion or (iii) nucleation. While the former two experimentally verified in nanosized magnetic tunnel junctions [1-2] or magnetic stripes with Néel-type chiral domain walls [23-24] demonstrate sharp switching characteristics, the nucleation switching driven by SOT can probably provide continuous switching behavior but has been rarely reported unambiguously [25].

For a magnetic stripe, usually, tiny domains are initially born at one boundary once a large enough current is applied [26-27] and then the nucleated domains grow larger very fast with elevating the current amplitude further. In this case, the switching process driven by SOT is easily dominated by the domain wall motion mechanism. Here, by applying a focused ion beam illumination technology, we can define an area with reduced perpendicular magnetic anisotropy and lower coercivity in comparison with pinned spins at unilluminated edges of stripe. These dedicated-designed devices make nucleation from the inner illuminated regions feasible, which provides us an ideal platform to study the domain nucleation process activated by SOT as following. As-resulted continuous switching properties can well be explained by a domain nucleation model. This technology can guide the design of SOT devices with continuous switching properties and can be applied in the coming neuromorphic computing in a spintronic manner.

Polycrystalline films of the composition Pt(5)/Co(0.6)/MgO(2)/Pt(2) (the thickness of the layers is indicated in nm) were prepared by magnetron sputtering at room temperature (RT) on substrates of naturally oxidized silicon. The base pressure in the chamber was $1.33 \times 10^{-5}$ Pa. The pressure of argon ($P_{Ar}$) during the deposition of Co and Pt layers was 0.16 Pa, for MgO - 0.13 Pa. At a fixed power of 100 W for direct (DC) and alternating (RF) voltage sources, the deposition rates were $V_{Pt}$=0.34 nm/s, $V_{Co}$=0.15 nm/s and $V_{MgO}$=0.08~0.1 nm/s. Using FIB setup combined with a scanning electron microscope (SEM), a stripe structure 120 μm×30 μm with two contacts for current propagation was prepared (Fig.2a). Then FIB was used to illuminate four regions with size 15 μm×20 μm for local modification of magnetic properties. For irradiation a beam of Ga$^+$ were used with a corresponding accelerating voltage of 5 kV and a beam current of 0.15 nA. The radiation dose was 1, 2, 3 and $4\times10^{-13}$ C/μm$^2$, respectively. Exposure to the ion beam leads to both a decrease in PMA and a decrease in the magnetization of Co layer.

Initially, the deposited film has the saturation magnetization $M_S$=1.23×10$^6$ A/m, the effective magnetic anisotropy energy $K_{eff}$=2.2×10$^5$ J/m$^3$ (with anisotropy field $H_K$ = 3.6 kOe measured by a vibrating sample magnetometer) and coercive force $H_C$=606 Oe. It is difficult to estimate the magnetic properties of the irradiated regions because of their small size. Indirect characteristics indicate that the saturation magnetization decreases, presumably due to mixing at the MgO/Co interface and implantation of Ga$^+$ ions. Considering the process of demagnetization for different illuminated regions by Kerr microscopy under the action of the in-plane field $H_x$, we can estimate the decreasing of $H_K$ to 2.5-3 kOe and the $H_C$ reduction to 230-280 Oe after ion exposure.

The SOT-measurements of the sample was performed together with a magneto-optical microscope based on the Kerr effect. The action was carried out by passing current pulses with a duration of 1 ms in the presence of a field in the plane ($H_x$). The temperature of the sample holder ($T_{ext}$) was controlled and measured during the experiment. In this experiment, a resistive heater was used to change $T_{ext}$. For each set of field and temperature values, the sample was remagnetized by a sequence of 20 current pulses with growing amplitude. After each pulse, a snapshot of the magnetic structure was taken.

Before going deep into the experimental data, we first introduce our domain nucleation model to understand the continuous nucleation switching process activated by SOT in the following. The illuminated region has an effective PMA of $K$ and saturation magnetization of $M_S$. $M_S$ is set as 1 in the model for simplicity. Before domain nucleation, the region can still be regarded uniformly magnetized and the magnetization **m** can be titled by SOT torque **τ** as described by the equation **m**×**H**$_{eff}$+τ**m**×**σ**×**m**=0, or Equation (1a) after simplification [28-29], the solution of which determines the final steady states. Here **H**$_{eff}$ is determined by -∂$E$/∂**m** and the system energy $E$ consists of the effective anisotropy -$Km_z^2$ and Zeeman energy -$m_xH_x$ or $E$=-$Km_z^2$-$m_xH_x$. The spin-orbit torque $\tau=\hbar/(2eM_St)\theta_{SH}J$ with $t$ - thickness of the ferromagnetic layer, $\theta_{SH}$ - spin Hall angle of the heavy metal, and $J$ - current density flowing in the heavy metal. The symbol **σ** denotes spin polarization of the spin current. To solve Eq. (1a), we construct another function of so-called virtual energy $E_{virt}$ as Eq. (1b). Straightforwardly, the local minima of Eq.(1b) correspond to the solutions of Eq.(1a). Worth noting, the $E_{virt}$ is not a real energy of the system. It is constructed only for solving Eq.1(a).

$$K \sin 2\theta + M_S H_x \cos\theta + \tau = 0 \tag{1a}$$

$$E_{virt} = K \sin^2\theta + M_S H_x \sin\theta + \tau\theta \tag{1b}$$

The energy profile $E_{virt}$ as the function of the magnetization orientation $\theta$, the polar angle respective to the film normal, can be directly shown in Fig. 1(a). As $\tau=0$, there exist two energy valleys which correspond to the two steady states. As $\tau$ is applied, the slope appears on the virtual energy profile. For the coherent switching, a critical torque $\tau_C \sim K$ [29, 30] has to be applied to make one valley turn to a saddle point. However, $\tau_C$ can be significantly reduced in the incoherent switching process. As shown in Fig.1(a), after $\tau$ is applied, one valley becomes more energetically favorable than the other. Worth of noting, the energy benefit $\Delta E_B$ from a possible switching is calculated by $E(\theta_1)-E(\theta_2)$. The $\theta_1$ and $\theta_2$ are the local and global minima of Fig.1(a), respectively. $\Delta E_B$ increases with $\tau$ (Fig.1(b)). Meanwhile, $\Delta E_B$ can also be deemed as the reduced bulk energy when a domain nucleates. Reasonably, the energy harvested from a nucleated domain can be expressed as $-\pi r^2 \Delta E_B$. Here $r$ is the average radius of nucleating domains. On the other hand, the birth of a domain will inevitably increase the boundary energy by $2\pi r E_D$. $E_D$ is the domain wall energy per square meter. The total energy to form a domain becomes $-\pi r^2 \Delta E_B + 2\pi r E_D$. In this case, only reaching a critical radius $r_C = E_D/\Delta E_B$ can the nucleating domain eventually survive, which also leads to an energy barrier $E_C = \pi E_D^2/\Delta E_B$ to nucleate a domain. Intuitively, the increase in $\tau$ and $H_x$ leads to the increase in $\Delta E_B$ and then the decrease in $E_C$. In a nucleation process, this barrier $E_C$ can be overcome by thermal agitation $k_B T$. Therefore, physically, the reversed areas $\Delta m$ via domain nucleation at $\tau$ shall be proportional to the following three factors as described by Eq. (2), (i) the switching probability $\exp(-E_C/k_B T)$ activated by thermal energy, (ii) the remaining unreversed areas $(1-m)$, and (iii) the average size of newborn nucleating seeds. Here $k_0$ is a scaling factor relating to the attempt frequency.

$$\frac{dm}{d\tau} = k_0 \cdot \exp\left(-\frac{E_C}{k_B T}\right) \cdot (1-m) \cdot \pi r_C^2 \tag{2}$$

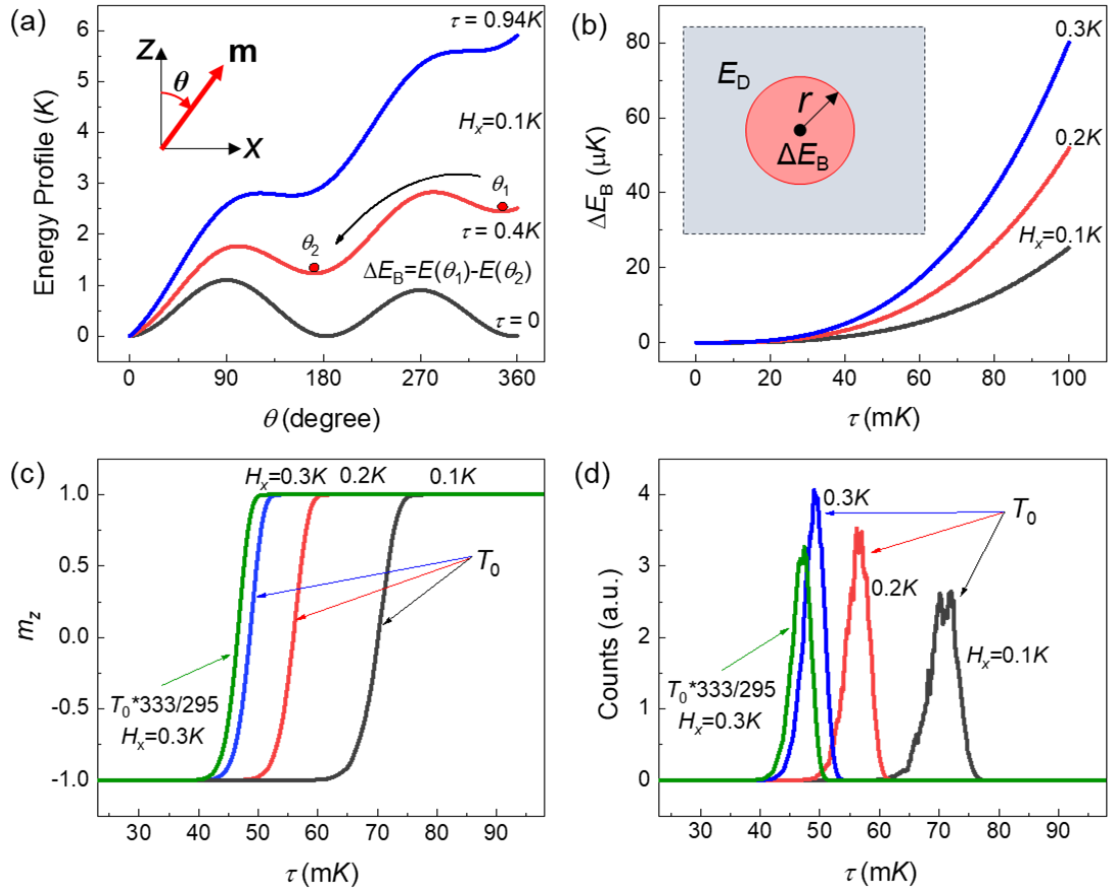

Fig. 1 (a) Energy profile as a function of magnetization orientation at different torques $\tau$ and $H_x=0.1K$. (b) Energy benefit $\Delta E_B$ due to nucleation of a domain at different $\tau$ and $H_x$. (c) Calculated SOT induced magnetization switching at different $H_x$. (d) Number of newborn domains as a function of $\tau$ at different $H_x$.

According to this nucleation model described by Eq. (1-2), the switching curves (half loop) can be numerically calculated as Fig.1(c). Worth noting, $\tau_C<0.1K$ is reduced by one order, compared with the coherent switching model which gives $\tau_C\sim K$. The model also gives out the other two results. (i) The switching probability is much more sensitive to $\tau$ than the critical radius $r_C$. Thus, the nucleation process is accomplished mainly by nucleating more domains instead of nucleating larger domains. We can then use $(dm/d\tau)/(\pi r_C^2)$ to estimate the number of newborn domains at $\tau$ (Fig. 1(d)). In the figure, $T_0=\pi E_D^2/k_B T$, reversely proportional to temperature and set as $10^{-4}K$ at room temperature. (ii) The switching probability is sensitive to $H_x$ and $T$. Thus, the increase in these parameters should noticeably reduce the critical switching current. In the following, we will experimentally verify the model by the FIB-defined devices.

The process of current-induced magnetization reversal was investigated for the selected region (embraced by a dashed rectangular) as an example in Fig.2(a). Fig.2(b-c) shows the visualization of the magnetization reversal process at $H_x$=0.3 or 0.9 kOe, respectively. Applied pulse current ($I$) is indicated in the left corner. With the increase in $I$, increasingly domains nucleate. Furthermore, a larger $H_x$ leads to finer but more domains at lower current, noticeably.

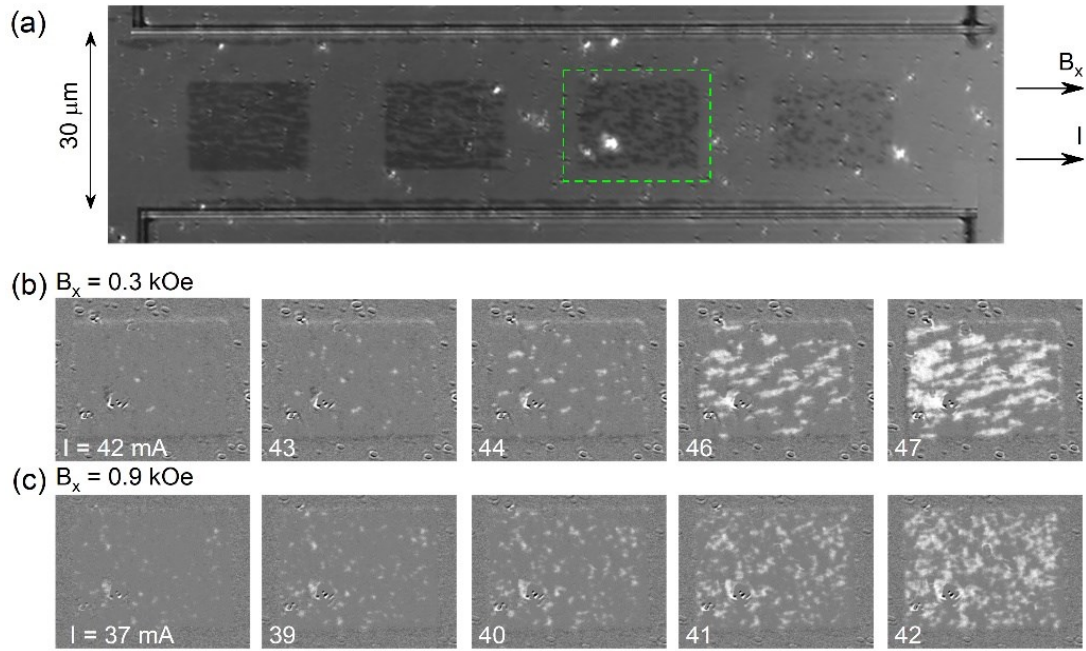

Fig.2. (a) Kerr microscope image of a Hall bar with four areas exposed to different doses of radiation (image without digital subtraction of background). The green rectangle marks one example. (b-c) Visualization of the process of magnetization reversal of selected areas under the action of current (current strength is indicated in the left corner) at $H_x$=0.3 kOe and 0.9 kOe, respectively (images taken with digital subtraction of the background for brighter display of domains).

Some domains also initially nucleate at the edges of the strip but not at the boundary between the FIB-illuminated and unilluminated regions, demonstrating that the spins at the boundary are pinned to some extent and nucleation can only be initialized inside the illuminated regions (not shown here). Data measured in other fields show similar results. We processed the obtained photos using the Image J program using automation with scripting [31]. The processing made it possible to obtain the value of the number of domains for each photograph.

Fig. 3(a) shows a typical measured magnetization switching curve ($M_z/M_S$) as a function of $I$, which seems similar with the calculated results in Fig. 1(c). Fig.3(b) gives the dependences of the number of domains on $I$ from the Kerr microscopy snapshots. It shows that the dependence has a peak as indicated in the theoretical results (see Fig.1(d)). The presence of a peak is due to the competition between two processes: the nucleation of domains and their union. It can be noted that an increase in the field $H_x$ leads to a significant increase in the maximum number of domains at a fixed temperature (Fig.3(d)), which is also reproduced in Fig.1(d). This feature results from the reason that a large $H_x$ increases $\Delta E_B$ and thus decreases the energy barrier $E_C$ to nucleate a domain. In this case, more seeds can nucleate once the critical current is reached.

An increase in $T_{ext}$ at a fixed field value has an ambiguous effect on the number of domains (Fig.3(b, d)). Our model predicts a declining trend of the domain count with elevating $T_{ext}$ (Fig.1(d)), which is interpretable as following. The domain number is mainly determined by the nucleation probability or its exponential fact $-E_C/k_BT=-\pi E_D^2/(\Delta E_B k_B T)$. A higher $T_{ext}$ reduces $I_C$ or $\tau_C$ (Fig.3(c)) and thus leads to a smaller $\Delta E_B$ (Fig.1(d)). The factor ($\Delta E_B k_B T$) can thus be even smaller with the increase in $T_{ext}$. This reason results in the decline in the domain number in Fig.1(d). However, practically, $E_D^2=AK$ [32] can also be influenced by temperature via decreasing the effective anisotropy $K$. Here $A$ is the exchange stiffness between neighboring spins. In this case, high temperature can also decrease the nucleation barrier and increase the domain number, which has not been taken into account in the model yet. The above two reasons may account for the ambiguous temperature dependence in Fig.3(d).

When comparing Fig.1(d) and Fig.3(b), we should point out their hidden relevance. The former calculates the number of newborn domains between $\tau$ and $\tau+d\tau$ at each $\tau$ while the latter calculates the total number of domains at $\tau$. By checking their definitions, it appears the two figures cannot be mutually mapped directly. However, we should point out in this experiment that those newborn small domains trend to merge with their neighbors to form a larger domain to thermodynamically reduce the domain wall energy of the system. After amalgamation driven by the domain wall energy relaxation, the newborn small seeds would take more weight in the total domain count than the old-born and merged domains with larger size. In this sense, taken domain union process into account, Fig. 3(b) tends to biasedly indicate the number of newborn domains. It is the reason why Fig.3(b) shares a similar line-shape with Fig.1(d).

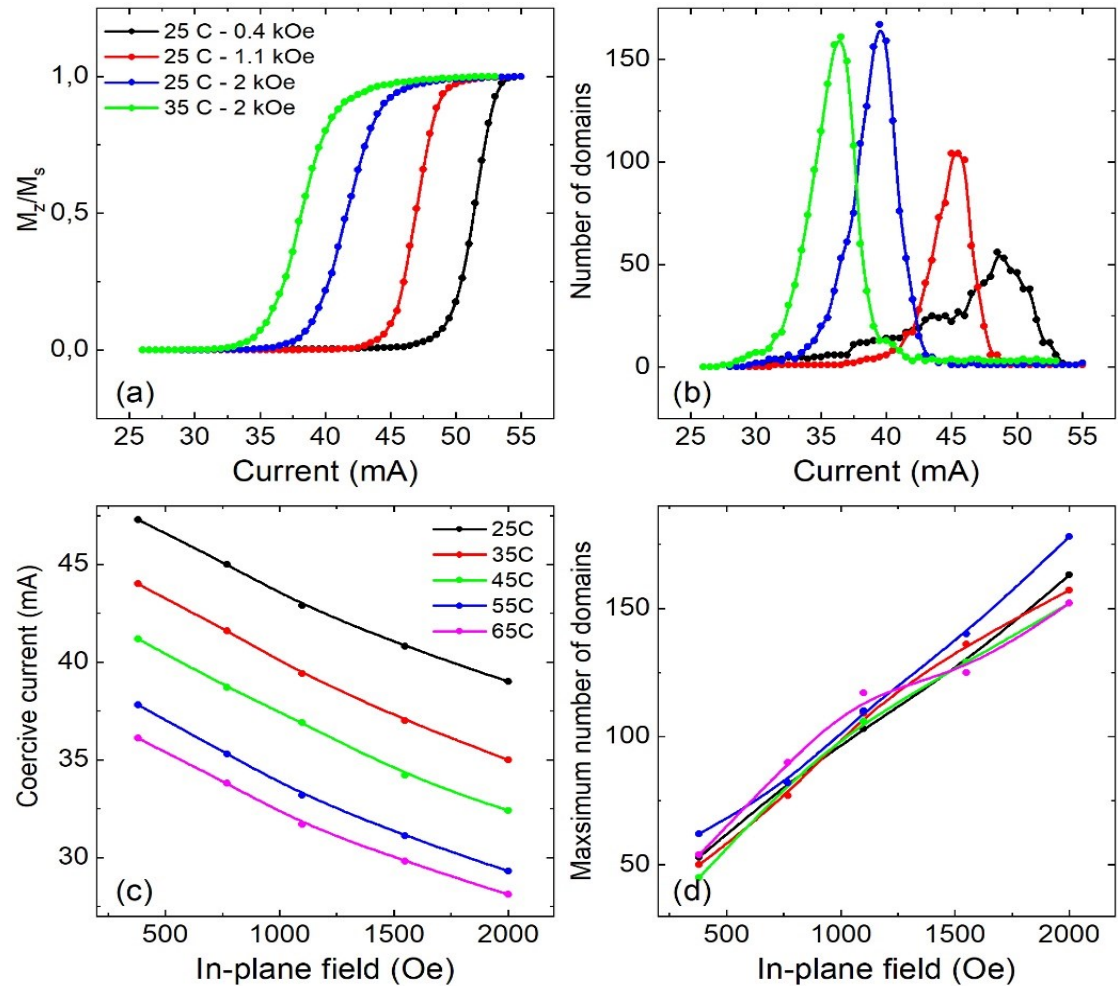

Fig. 3 (a) Half of the hysteresis loop of current-induced magnetization reversal for different $H_x$ and two values of $T_{ext}$. (b) Dependence of the number of domains in the considered region on the current value for different values of the field $H_x$ and two $T_{ext}$ values. (c) Plot of the current required for magnetization reversal on the magnitude of the field $H_x$ for different $T_{ext}$ values. The dependence of the nucleation current behaves in the same way. (d) Dependence of the maximum number of domains in the considered region on the magnitude of the field $H_x$ for different values of the $T_{ext}$.

Based on the obtained photographs, the dependences of the magnetization on the current amplitude (half hysteresis) were summarized in Fig.3(a). The other half is symmetrical to the shown half, thus not given for simplicity. It can be noted that the value of the current corresponding to the onset of the magnetization reversal process significantly decreases, both with an increase in $H_x$ and $T_{ext}$. The increased $H_x$ can enlarge the energy benefit $\Delta E_B$ and thus reduce the energy barrier $E_C$ for nucleation. On the other hand, higher $T_{ext}$ improves the nucleation probability by enhancing thermal agitation. Both factors make domain nucleation easier and thus the onset current is reduced. The experimental curves in Fig.3(a) therefore look similar to the curves reproduced by the nucleation model in Fig.1(c).

The critical switching current when $H_x$=1.1 kOe and $T_{ext}$=25 °C is about 44 mA which corresponds to 2.3×10$^7$A/cm$^2$ if supposing current density uniformly distributed in the metallic layers. If the coherent switching model is adopted, for our samples with $M_S$=1.23×10$^6$ A/m (an overestimated value), $t_{Co}$=0.6 nm, $H_K$=3 kOe and spin Hall angle of Pt $\theta_{SH}$=0.1, the calculated $J_C$ according to the coherent switching model [30] would be 3.4×10$^8$A/cm$^2$, about one order higher than the observed one. In Fig.1(c) the critical torque $\tau_C$ is about $0.07K$ while $\tau_C \sim K$ in the coherent model, also reduced by a factor of 14. Here, clearly, the nucleation mechanism is very helpful to make SOT switching happen in a one-order-lowered $J_C$. Most importantly, magnetization switching is accomplished by nucleation of hundreds of domains with small sizes, which smoothens the switching curves. Thus, those intermediate states become accessible by sourcing the proper current. Noticeably, a larger $H_x$ inclines to flatten the switching curve and make the switching gentler as shown in Fig.3(a). Here we use two standards to evaluate the flatness of the switching curve in Fig.3(a). One is the maximum slope which is 0.3 and 0.17/mA for $H_x$=0.4 and 2 kOe, respectively. The other is the current range in which magnetization is switched from 20% to 80% saturated magnetization. This current range is about 2.2 and 3.5 mA for $H_x$=0.4 and 2 kOe, respectively. Both standards show the higher $H_x$ makes the switching gentler. It is probably because the large $H_x$ can reduce the energy barrier to nucleate domains and thereafter the number of domains (or the number of intermediate states) are remarkably increased as shown in Fig.1(d) and Fig.3(b). This advantage can be made full use in neuromorphic computing applications where continuous switching is preferred.

In conclusion, we have unambiguously observed continuous nucleation switching driven by SOT in ion-beam defined devices. Experimental study of the switching process by Kerr microscopy and electrotransport measurements confirmed the results of the proposed theoretical model. The number of newborn domains shows a peak feature as the function of the applied SOT, which can be further tuned by an applied $H_x$. A larger $H_x$ increases the bulk energy benefit from domain nucleation, reduces the energy barrier for domain nucleation, and thus increases the number of domains formed during the switching process. The experimental and simulated results reproduced each other reasonably well, including the current dependence of the magnetization, the critical current density, and the current dependence of the number of newborn domains. This nucleation switching mechanism is helpful to realize gentle SOT-switching with intermediate states easily accessible, which can be potentially utilized in neuromorphic computing.

**Acknowledgements**:
X. F. Han and C. H. Wan appreciates the financial support from the National Key Research and Development Program of China (MOST) (Grant No. 2017YFA0206200), the National Natural Science Foundation of China (NSFC) [Grant No. 51831012, 11974398, 51620105004, 11811530077], Beijing Natural Science Foundation (Grant No. Z201100004220006), the Strategic Priority Research Program (B), the Key Research Program of Frontier Sciences and the International Partnership Program of Chinese Academy of Sciences (CAS) [Grant No. XDB33000000, QYZDJ-SSW-SLH016, and 112111KYSB20170090]. C. H. Wan appreciates financial support from Youth Innovation Promotion Association, CAS (2020008). A. S. Samardak acknowledge the Russian Ministry of Education and Science under the state task (0657-2020-0013) and Act 211 of the Government of the Russian Federation (Contract No. 02.A03.21.0011). A. V. Ognev thanks the program of improvement of the competitiveness of Far Eastern Federal University (Agreement No. 075-02-2020-1584) and the Russian Foundation for Basic Research (Grant No. 1902-00530). A. G. Kolesnikov thanks the scholarship of President of the Russian Federation for young scientists and graduate students (SP-350.2019.1).

**Availability of data**: The data that support the findings of this study are available from the corresponding author upon reasonable request.